\documentstyle[preprint,prc,eqsecnum,aps,epsf]{revtex}

\tightenlines   

\begin{document}
\draft

\def\lsim{\thinspace{\hbox to 8pt{\raise -5pt\hbox{$\sim$}\hss{$<$}}}\thinspace}
\def\rsim{\thinspace{\hbox to 8pt{\raise -5pt\hbox{$\sim$}\hss{$>$}}}\thinspace}

\title{Energy Dependence of the NN t-matrix in the Optical Potential
for Elastic Nucleon-Nucleus Scattering} 

\author{ Ch.~Elster, S.P.~Weppner}
\address{
Institute of Nuclear and Particle Physics,  and
Department of Physics, \\ Ohio University, Athens, OH 45701}


\date{\today}

\maketitle
 
\begin{abstract}
The influence of the energy dependence of the free NN t-matrix on the
optical potential of nucleon-nucleus elastic scattering is investigated
within the context of a full-folding model based on the impulse
approximation. The treatment of the pole structure of the NN t-matrix,
which has to be taken into account when integrating to negative energies
is described in detail. We calculate proton-nucleus elastic scattering
observables for $^{16}$O, $^{40}$Ca, and $^{208}$Pb between 65 and 200~MeV 
laboratory energy and study the effect of the energy dependence of
the NN t-matrix. We compare this result with experiment and with
calculations where the center-of-mass energy of the NN t-matrix is
fixed at half the projectile energy. It is found that around 200 MeV the
fixed energy approximation is a very good representation of the full
calculation, however deviations occur when going to lower energies (65~MeV).
\end{abstract}


\pagebreak


\narrowtext


\section{Introduction}

\hspace*{10mm}
The scattering of protons and neutrons from nuclei has a long
history as a  tool to investigate the details of the reaction mechanism
between a nucleon and a many nucleon system. The spectator
expansion of multiple scattering theory \cite{Corr,Sicil,med2} is our
theoretical approach to define an optical potential for elastic 
nucleon-nucleus scattering. This expansion is predicated upon the idea
that two-body interactions between projectile and target nucleons play
the dominant role.

\hspace*{10mm}
In its most general form, the first order single scattering optical
potential within the framework of the spectator expansion is given by
the expectation value of the nucleon-nucleon (NN) transition 
amplitude and the ground state of the target nucleus. This
`full-folding' optical potential involves the convolution of
the fully-off-shell two nucleon scattering amplitude with a realistic
nuclear density matrix. In this form, the exact calculation of this 
optical potential requires a three-dimensional integration, in which
the integration variable is coupled to the energy of propagation 
of the projectile and target nucleon. This very fact leads to 
a full-folding optical potential, which explicitly treats the 
off-shell behavior {\it and} the energy dependence of the NN t-matrix
when carrying out the integration. Full-folding models along this
line have been proposed and carried out \cite{hugo1,hugo2} with
the conclusion that these model calculations based on the free
NN t-matrix give a good description of the data for energies 
between 200 and 400~MeV, however are rather poor  at 200~MeV and
below. We have repeated these calculations within our computational
framework based on a full-folding model employing a realistic
nuclear density matrix and a free NN t-matrix where the off-shell
behavior as well as the energy dependence are taken into account. 
However, we do not find 
as large effects as shown in Refs. \cite{hugo1,hugo2}.

\hspace*{10mm}
It can be argued that at intermediate energies the scattering
of a projectile from a nucleon in the target nucleus may
resemble free NN scattering mainly in forward direction. This is
the justification for a common approximation to the optical potential
in which the energy of the NN t-matrix is uncoupled from the integration
variable by fixing it at half the projectile laboratory energy. We will
study the accuracy of this assumption at different projectile energies.

\hspace*{10mm}
The structure of this paper is as follows. First we will review in
Section II the
relevant expressions for the single scattering optical potential
in the impulse approximation as well as the full-folding procedure
as used in our calculations. We will describe in some detail
the numerical implementation for treating the energy dependence of
the NN t-matrix, especially since this was left out in Ref. \cite{hugo1}.
We then present in Section III
elastic scattering results for proton scattering
from a variety of nuclei in the energy regime between 65 and 200~MeV,
and end with concluding remarks in Section IV.

\section{Theoretical Formulation}

\hspace*{10mm}
The transition amplitude for elastic scattering of a projectile from
a target nucleus is given as \cite{med2}
\begin{equation}
T_{el} = P U P + P U G_0(E) T_{el}, \label{eq:2.1}
\end{equation} 
where $P$ is the projector on the ground state $|\Phi_A\rangle$ of
the target, $P = \frac{|\Phi_A\rangle \langle \Phi_A|}
{\langle \Phi_A| \Phi_A\rangle}$, and $G_0(E)=(E-H_0 +i\varepsilon)^{-1}$.
For the scattering of a single particle projectile from an A-particle
target nucleus the free Hamiltonian is given by $H_0=h_0+H_A$, where $H_A$ 
stands for the target Hamiltonian.  In the spirit of the spectator
expansion the target Hamiltonian is viewed as $H_A=h_i + \sum_{j\neq i}
v_{ij} +H^i$, where $h_i$ is the kinetic energy operator for the
$i$th target nucleon,  $v_{ij}$  the interaction between target
nucleon $i$ and the other target nucleons $j$, and $H^i$ is an 
(A-1)-body operator containing all higher order effects.
In a mean field approximation $\sum_{j\neq i} v_{ij} \approx W_i$, 
where $W_i$ is assumed to depend only on the $i$th particle
coordinate. Thus, the propagator consistent with the first order
in the spectator expansions is given as
\begin{equation}
G_0(E) \approx G_i(E)=[(E-E^i)-h_0 -h_i -W_i + i \varepsilon]^{-1}.
 \label{eq:2.2}
\end{equation} 
Here $H^i$, having no explicit dependence on the $i$th particle,
is replaced by an average
energy $E^i$, which is at most equivalent to the separation energy of a 
nucleon from the nucleus. In most of our calculations we set $E^i=0$,
since the projectile energies are in comparison much larger. 
We will test this assumption. Since we are considering the optical
potential in the impulse approximation, we neglect the mean field
$W_i$ of Eq.~(\ref{eq:2.2}) in the following.

\hspace*{10mm}
The driving term of Eq.~(\ref{eq:2.1}) denotes the optical potential,
which in first order is given as
\begin{equation}
\langle {\bf k'}|\langle \Phi_A| PUP|\Phi_A\rangle |{\bf k}\rangle
\equiv {\hat U}({\bf k'},{\bf k}) = \sum_{i=n,p}\langle {\bf k'}| 
\langle \Phi_A| {\hat \tau}_{0i}({\cal E})|\Phi_A \rangle |{\bf k}\rangle.
  \label{eq:2.3}
\end{equation}
Here ${\bf k'}$ and ${\bf k}$ are the external momenta of the system,
${\hat \tau}_{0i}({\cal E})$ represents the NN transition operator
\begin{equation}
{\hat \tau}_{0i}(E) = v_{0i} + v_{0i} g_i(E) {\hat \tau}_{0i} ,
\label{eq:2.4}
\end{equation}
with
\begin{equation}
g_i(E)= [(E-E^i) -h_0 -h_i + i \varepsilon]^{-1}, \label{eq:2.5}
\end{equation}
and $v_{0i}$ representing the NN interaction. The sum over $i$ in
Eq.~({\ref{eq:2.3}) indicates the two different cases, namely when the
target nucleon is one of Z protons, and when it is one of N neutrons.
The energy ${\cal E}$ is the energy of the interacting system. Inserting
a complete set of momenta for the struck target nucleon before and
after the collision and evaluating the momentum conserving $\delta$-functions
gives as final expression for the full-folding optical potential 
\cite{ff1,swth}
\begin{eqnarray}
\hat{U}({\bf q}, {\bf K})= \sum_{i=n,p} \int d^3P &&
\;\eta({\bf P},{\bf q}, {\bf K})\;\hat{\tau}_{0i}({\bf q}, 
\frac{1}{2}(\frac{A+1}{A}{\bf K}-{\bf P}), {\cal E}) \nonumber \\
&&\rho_i({\bf P}-\frac{A-1}{A}\frac{{\bf q}}{2},
{\bf P}+\frac{A-1}{A}\frac{{\bf q}}{2}) \label{eq:2.7}.
\end{eqnarray}
Here the arguments of the NN amplitude $\hat{\tau}_{0i}$ are 
${\bf q}={\bf k}'-{\bf k}={\bf\cal K}'-{\bf \cal K}$ and 
$\frac{1}{2}({\bf \cal K}'+{\bf\cal K}) =
\frac{1}{2}(\frac{A+1}{A}{\bf K}-{\bf P})$, where
${\bf\cal K}'=\frac{1}{2}({\bf k}' - ({\bf P}-{\frac{\bf q}{2}} -
\frac{\bf K}{A})$ and 
${\bf \cal K}=\frac{1}{2}({\bf k} - ({\bf P}+{\frac{\bf q}{2}} -
\frac{\bf K}{A})$ are the nonrelativistic final and initial
nuclear momentum in the zero momentum frame of the NN system,
and ${\bf K}=\frac{1}{2}({\bf k'}+{\bf k})$.
The factor $\eta({\bf P},{\bf q}, {\bf K})$ is the M\o ller
factor for the frame transformation~\cite{Joachain}, and 
$\rho_i$ represents the density matrix of the target.
Evaluating the propagator $g_i(E)$ of Eq.~(\ref{eq:2.5}) in 
the nucleon-nucleus (NA) center of mass frame yields for the 
energy argument ${\cal E}$ of the NN amplitude 
$\hat{\tau}_{0i}$ of Eq.~(\ref{eq:2.7})
\begin{equation}
{\cal E} = E_{NA} - \frac{(\frac{A-1}{A}{\bf K}+{\bf P})^2}{4m_N}.
\end{equation}
Here $E_{NA}$ is the total energy in the NA center of mass frame
and $m_N$ is the nucleon mass. At this point we assume $E^i=0$ as
was discussed earlier. Since we employ relativistic definitions
for the NA kinematics we have $E_{NA}=
\sqrt{({\bf k}c)^2 +{(m_Nc^2)}^2}+\sqrt{({-\bf k}c)^2+{(m_Ac^2)}^2}$,
where $m_A$ denotes the mass of the nucleus.
With these definitions, the optical potential of Eq.~(\ref{eq:2.7})
becomes
\begin{eqnarray}
\hat{U}({\bf q}, {\bf K})= \sum_{i=n,p} \int d^3P &&
\eta({\bf P},{\bf q}, {\bf K})\;\hat{\tau}_{0i}({\bf q},
\frac{1}{2}(\frac{A+1}{A}{\bf K}-{\bf P}), 
E_{NA} - \frac{(\frac{A-1}{A}{\bf K}+{\bf P})^2}{4m_N}) \nonumber \\
&&\rho_i({\bf P}-\frac{A-1}{A}\frac{{\bf q}}{2},
{\bf P}+\frac{A-1}{A}\frac{{\bf q}}{2}) \label{eq:2.9}.
\end{eqnarray}
This  expression shows that the evaluation of the full-folding
integral requires the NN t-matrix not only off-shell but also
at energies $E_{NA} \geq {\cal E} > -\infty$. Specifically
when going to negative energies, we have to take into account the 
pole structure of the NN t-matrix. The NN interaction supports a 
bound state at $E_d=-2.225$ MeV in the $^3S_1-^3D_1$ channel, the deuteron,
and a virtual state in the $^1S_0$ channel at -66 keV, the `di-proton'.
A virtual state means a pole of the NN t-matrix in the 
second sheet of the complex energy plane, which manifests itself on the 
real axis as a very narrow finite peak (about 100 keV wide) around
the pole position.
The deuteron, a true bound state, causes a pole in the NN t-matrix at 
the binding energy $E_d$. In order to explicitly treat this pole, 
we factorize $\hat{\tau}_{0i}$ into a pole term and a residue function
\begin{equation}
\bar{\tau}_{0i}= \hat{\tau}_{0i}\times(
E_{NA} - \frac{(\frac{A-1}{A}{\bf K}+{\bf P})^2}{4m_N} - E_d),
\end{equation}
so that the optical potential takes the form
\begin{eqnarray} 
\hat{U}({\bf q}, {\bf K})= \sum_{i=n,p} \int d^3P && 
\;\frac{\bar{\tau}_{0i}({\bf q}, 
\frac{1}{2}(\frac{A+1}{A}{\bf K}-{\bf P}), 
E_{NA} - \frac{(\frac{A-1}{A}{\bf K}+{\bf P})^2}{4m_N})}
{E_{NA} - \frac{(\frac{A-1}{A}{\bf K}+{\bf P})^2}
{4m_N} - E_d +i\epsilon}
 \nonumber \\
&&\eta({\bf P},{\bf q}, {\bf K})
\rho_i({\bf P}-\frac{A-1}{A}\frac{{\bf q}}{2}, 
{\bf P}+\frac{A-1}{A}\frac{{\bf q}}{2}) \label{eq:2.11}.  
\end{eqnarray} 
To obtain a simple pole, which can be treated with standard numerical
methods, we perform a change of the integration variable to 
${\bf Q}=\frac{A-1}{A}{\bf K}+{\bf P}$, and Eq.~(\ref{eq:2.11}) becomes
\begin{eqnarray}  
\hat{U}({\bf q}, {\bf K})= \sum_{i=n,p} \int d\hat{Q} && \int_0^\infty 
\;dQ Q^2\; 4 m_N\;\frac{\bar{\tau}_{0i}({\bf q},  
\frac{1}{2}({2\bf K}-{\bf Q}),  
E_{NA} - \frac{Q^2}{4m_N})}
{Q_d^2-Q^2+i\epsilon}
 \nonumber \\
&&\eta({\bf Q},{\bf q}, {\bf K}) 
\rho_i({\bf Q}-\frac{A-1}{A}({\bf K}+\frac{{\bf q}}{2}),   
{\bf Q}-\frac{A-1}{A}({\bf K}-\frac{{\bf q}}{2})) \nonumber \\
=\sum_{i=n,p} \int d\hat{Q} \;&& {\cal P}\int_0^\infty
\;dQ Q^2\; 4 m_N\;\frac{\bar{\tau}_{0i}({\bf q},
\frac{1}{2}({2\bf K}-{\bf Q}),
E_{NA} - \frac{Q^2}{4m_N})}
{Q_d^2-Q^2}
 \nonumber \\
&&\eta({\bf Q},{\bf q}, {\bf K})
\rho_i({\bf Q}-\frac{A-1}{A}({\bf K}+\frac{{\bf q}}{2}),
{\bf Q}-\frac{A-1}{A}({\bf K}-\frac{{\bf q}}{2})) \nonumber \\
-i \sum_{i=n,p} \int d\hat{Q} \; &&2\pi m_N Q_d\;
{\bar{\tau}_{0i}({\bf q},
\frac{1}{2}({2\bf K}- Q_d\hat{Q}), E_d)} \nonumber \\
&&\eta(Q_d\hat{Q},{\bf q}, {\bf K}) 
\rho_i(Q_d\hat{Q}-\frac{A-1}{A}({\bf K}+\frac{{\bf q}}{2}),
Q_d\hat{Q}-\frac{A-1}{A}({\bf K}-\frac{{\bf q}}{2}))
\label{eq:2.12}.  
\end{eqnarray} 
Here $\int d\hat{Q}$ represents the angular integration, 
$Q_d=\sqrt{4m_N(E_{NA}-E_d)}$, and ${\cal P}$ denotes a 
Cauchy principal value integral.

\hspace*{10mm}
The cut in the integrand of Eq.~(\ref{eq:2.12}) indicates the opening of 
a new channel in the scattering reaction. In this specific case
it is the deuteron pickup channel, describing when an incoming 
projectile proton picks up a neutron and a deuteron is knocked out,
thus removing flux from the elastic channel. In order to obtain some 
a priori estimate on the possible size of this additional channel
on the elastic
scattering reaction, it is worthwhile to look at experimental information
on the deuteron pickup reaction.

\hspace*{10mm}
In the 1980's cross sections of the (p,d) reaction have been measured
for the closed shell nuclei considered here~\cite{Hosono,Abegg,Kraushaar}.
For $^{40}$Ca at 200 MeV, the differential cross section for the 
(p,d) reaction for small angles (10-20 degrees) is about 
1 mb/sr while at 65 MeV it is about 10 mb/sr. Compared to the typical 
size of differential cross sections for elastic scattering in a 
similar angular range (cp. Figs.~2 and 6), this is quite small.
The effect of this additional channel will be studied in our calculations
of elastic scattering observables and compared to calculations where
the integration variable is decoupled from the energy of propagation
of the projectile and target target nucleon. The latter is an 
approximation where the energy of the NN t-matrix is fixed at half
the beam energy in the laboratory frame
\begin{equation}
{\cal E} = E_0 = \frac{1}{2} \frac{k^2_{lab}}{2m_N}=
\frac{1}{2}\frac{{(\frac{A+1}{A}k_0)}^2}{2m_N}\label{eq:2.12b},
\end{equation}
where $k_{lab}$ and $k_0$ are the on-shell momenta in the 
laboratory and NA system respectively. With this fixed energy approximation
the full-folding optical potential of Eq.~(\ref{eq:2.11}) becomes
\begin{eqnarray}
\hat{U}({\bf q}, {\bf K})= \sum_{i=n,p} \int d^3P &&
\eta({\bf P},{\bf q}, {\bf K})\;\hat{\tau}_{0i}({\bf q},
\frac{1}{2}(\frac{A+1}{A}{\bf K}-{\bf P}),
E_0) \nonumber \\
&&\rho_i({\bf P}-\frac{A-1}{A}\frac{{\bf q}}{2},
{\bf P}+\frac{A-1}{A}\frac{{\bf q}}{2}) \label{eq:2.13}.
\end{eqnarray}
Full-folding calculations based on this approximation have been
carried out by several groups \cite{ff1,FFC}.

\hspace*{10mm}
Fixing the energy at $E_0$ as given in Eq.~({\ref{eq:2.12b}}) is a
`historic' choice, made by all calculations based on the KMT 
formulation~\cite{KMT}.
The argument for this specific choice of $E_0$ is, that this is the
energy at which free NN scattering at a fixed target nucleon would
occur.  This choice of $E_0$ favors the forward scattering
process, and it was argued that due to the narrow 
peaking of the density in 
momentum space, the scattering is dominated by
forward scattering. Having in mind that the differential cross section
for e.g. proton scattering from $^{208}$Pb at 200 MeV falls off by
five orders of magnitude between 5 and 30 degrees,
this argument may capture some truth, however  needs to be tested 
numerically.

\hspace*{10mm}
Before discussing the results of our calculations, we would like to 
elaborate some more on the details of our numerical implementation.
Treating the pole structure of the NN t matrix while integrating over the
energy of the propagation is standard in modern three-nucleon scattering
calculations~\cite{witala}. However, these calculations are carried out
in partial waves, which allows treating the $^3S_1-^3D_1$ and 
$^1S_0$ channel separately. Our calculation of the full-folding 
optical potential of Eq.~(\ref{eq:2.12}) is carried out directly in 
three dimensions based on Monte Carlo integration. Thus, for the
neutron-proton part of the optical potential, we have to treat the 
pole singularity and the virtual state simultaneously. The principle
value integral of Eq.~(\ref{eq:2.12}) is treated with standard
subtraction techniques. When calculating the optical potential as given in 
Eq.~(\ref{eq:2.12}), we have to interpolate the NN t-matrix in four
dimensions, $|{\bf q}|, |2{\bf K}-{\bf Q}|, (2{\bf K}-{\bf Q})\cdot
{\bf q}$, and the energy. For the momenta we use a three dimensional
B-spline, for the energy a linear interpolation, since over a large
range of energies the NN t-matrix
is a slowly varying function of the energy. When carrying out the energy
integration of Eq.~({\ref{eq:2.12}), we find that for the higher
projectile energies integrating out to -100~MeV c.m. energy in the 
NN t-matrix is sufficient,
while for the lower energies (65 MeV) we need to integrate
out to -400 MeV c.m. energy.

\hspace*{10mm}
Of course, the argument of the NN t-matrix being a slowly 
varying function of the energy is not true in the 
immediate vicinity of the virtual state
in the $^1S_0$ channel. Since the peak around the pole position in the 
second energy sheet is finite and very narrow, we evaluate 
the integral over the t-matrix in this region separately and define
\begin{equation}
\hat{\tau}_{av,i}({\bf q}, {\bf K}, E_{NA})= 
\frac{1}{\Delta_1+\Delta_2}
\int^{Q_0+\Delta_2}_{Q_0-\Delta_1}dQ \hat{\tau}_{0i}({\bf q},
\frac{2\bf K - \bf Q}{2},E_{NA}-\frac{Q^2}{4m_N}),\label{eq:2.14}
\end{equation}
where $Q_0$ is defined via $E_{NA}-\frac{Q_0^2}{4m_N}=0$, which is
close to 
the pole position on the second sheet of the complex energy plane.
The `average' $\hat{\tau}_{av,i}$ is usually obtained from 40 energy
points over a momentum interval of 0.5~MeV, and is then used as one of 
the interpolation points in the energy interpolation.
The midpoint $Q_0$ for the
integration in Eq.~({\ref{eq:2.14}}) depends via $E_{NA}$  on the 
projectile energy, which implies that the averaging process has 
to be carried out for each projectile energy. If we want to simplify the 
numerical procedure and create  $\hat{\tau}_{av,i}$ only once, we change
the integration variable from $Q$ to $E$ and calculate
\begin{equation} 
\hat{\tau}_{av,i}({\bf q}, {\bf K}, E_{0})= 
\frac{1}{2\delta_1}
\int^{E_0+\delta_1}_{E_0-\delta_1}dE \hat{\tau}_{0i}({\bf q},
\frac{2\bf K - \bf Q}{2},E)\label{eq:2.15},
\end{equation}
where we have chosen $E_0=0$ MeV and $\delta_1=0.25$ MeV. The error
in obtaining $\hat{\tau}_{av,i}$ from Eq.~({\ref{eq:2.15}}) instead
of Eq.~Eq.~({\ref{eq:2.14}}) is small if $\frac{2\delta_1}{E_{NA}}$
is small. For our worst case tested, $E_{NA} = 50$ MeV for proton scattering
from $^{208}$Pb the numerical error was 1\%.

\section{Results and Discussion}

\hspace*{10mm}
In this paper the study of elastic scattering of protons
from  spin-zero target nuclei at energies that range
from 65 to 200 MeV incident projectile energy is strictly first order
in the spectator expansion and based on the impulse approximation. The
full-folding optical potential is  calculated as outlined in the 
previous section, specifically as given in Eq.~(\ref{eq:2.12}).
As a model for the density matrix for the target
nucleus we employ a  Dirac-Hartree
calculation~\cite{DH}. The Fourier transform of the vector density,
$\rho({\bf r}',{\bf r})$, serves as our
non-relativistic single particle density~\cite{ff1}.
Another crucial ingredient in the calculation of 
the optical potential, $\hat{U}({\bf q}, {\bf K})$, is the fully
off-shell NN t-matrix. The calculations presented here employ the
NN t-matrix based on the  charge-dependent Bonn potential \cite{CDBonn}. 
This potential is fitted to describe the 
Nijmegen database with a $\chi^2$ per datum $\sim$1. It is 
also to be understood that we perform all spin summations in 
obtaining $\hat{U}({\bf q}, {\bf K})$. This reduces the required NN
t-matrix elements to a spin independent component (corresponding
to the Wolfenstein amplitude A) and a spin-orbit component
(corresponding to the Wolfenstein amplitude C). Since we are
assuming that we have spin saturated nuclei, the components of the 
NN t-matrix depending on the spin of the struck nucleon vanish.
The Coulomb interaction between the projectile and the target
is included using the exact formulation as described 
in Ref.~\cite{coul}.

\hspace*{10mm}
At first we want to concentrate on proton scattering from different
target nuclei at an intermediate energy. In Fig.~1 we 
display the differential cross section $\frac{d\sigma}{d\Omega}$, 
the analyzing power $A_y$, and the spin rotation function $Q$, for
elastic proton scattering from $^{16}$O. The solid line 
represents the full calculation of the optical potential according 
to Eq.~(\ref{eq:2.12}) and the dashed line shows the calculation
where the energy of the NN t-matrix is fixed at half the projectile
energy. It is remarkable how close the fixed energy result is to the
full calculation. This may stem from a relatively weak energy
dependence of the NN t-matrix and the fact that the folding with
the sharply peaked density matrix does not sample the NN t-matrix in the
negative energy region. We confirmed this numerically by artificially
limiting the energy integration to positive energies and did not
find noticeable effects in the scattering observables at 200 MeV.
In Fig.~2 we show the observables for elastic proton scattering 
from $^{40}$Ca at 200 MeV and in Fig.~3  those for 
elastic proton scattering from $^{208}$Pb. As in Fig.~1, the solid line
represents the full calculation of the optical potential
taking the energy dependence of the NN t-matrix into account, 
whereas the dashed line shows the calculation at fixed energy
$E_0$ given in Eq.~(\ref{eq:2.13}). For all three nuclei the fixed
energy result is remarkably close to the full calculation, 
a conclusion which essentially was also drawn in Ref.~\cite{hugo1}.
A general trend is that for the fixed energy calculation the 
dip structure in the differential cross section and the spin
observables is slightly more pronounced. 

\hspace*{10mm}
In order to assess the
importance of the additional energy shift given by $E^i$ in
Eq.~(\ref{eq:2.2}), we show in Fig.~1 a calculation (including the
energy dependence of the NN t-matrix), which was performed setting
$E^i=-8$~MeV. This results in a shift of the total energy of the NA system
to a slightly higher value, as suggested already in Refs.
\cite{Crespo,med1}. The effect of this energy shift is negligible at
200~MeV as shown in Fig.~1.

\hspace*{10mm}
Next we turn to lower projectile energies, where we may expect to see
differences between the full calculation including the energy dependence
of the NN t-matrix and a calculation with a NN t-matrix energy fixed at
half the projectile energy. In Fig.~4 we display the results for proton
scattering from $^{208}$Pb at 160~MeV and in Fig.~5 those for proton
scattering from $^{16}$O at 135~MeV. Compared to the results at 200~MeV,
the difference between the solid line, which represents the full
calculation, and the dashed line, which represents the calculation for a
fixed energy of the NN t-matrix, becomes more pronounced, especially for
$A_y$ at larger angles as shown in Fig.~4. A similar figure is shown in
Ref.~\cite{hugo1}, however a direct comparison between calculations is
not possible, since different NN t-matrices as well as densities are
employed. 
The scattering observables given in Fig.~5 for proton scattering from
$^{16}$O at 135~MeV exhibit less difference between the solid and dashed
lines, leading to the conclusion that even at this relatively 
low energy the fixed energy prescription in the optical potential is
still amazingly and perhaps unexpectedly good. One common trend is
becoming apparent in Figs.~4 and 5, namely that the differential cross
sections predicted by the full calculations are systematically higher
compared to the ones predicted by the calculations employing a NN
t-matrix at a fixed energy. In Fig.~5 we also include a full
calculation, in which the energy shift $E^i$ is chosen to be 
$E^i=-8$~MeV (dash-dotted line). Again, we conclude that the effect of
this energy shift is minute.

\hspace*{10mm}
Last we want to consider scattering observables at an energy below
100~MeV projectile energy. It has already been stated in the literature
\cite{med2,hugo2,hugo3} that at these low energies the strict impulse
approximation is insufficient to describe the experimental observables.
We arrive again at the same conclusion, when showing in Fig.~6
the scattering observables for proton scattering from $^{40}$Ca at
65~MeV.  However, this is not the main point we want to make. When
comparing a calculation including the energy dependence of the NN
t-matrix in the optical potential (solid line) with a calculation where
the energy is fixed at half the projectile energy (dashed line), we
clearly see that the additional channels, the deuteron pick-up channel
and the di-proton state have a sizable effect on the elastic
observables. This is consistent with the experimental data, which show
that the cross section for the (p,d) reaction 
is getting larger at lower energies.
The differential cross section predicted by the full calculation (solid
line) is
higher than the one predicted by the calculation using the fixed energy
prescription (dashed line), 
and actually slightly closer to the experiment. The spin
observables obtained from the full calculations show much less structure
compared to those from the fixed energy calculation. However, both
impulse approximation calculations do not adequately describe the spin
observables. A similar conclusion was drawn in Ref.~\cite{hugo2},
although there the effect of including the deuteron and di-proton
channel is more dramatic than in our calculations. For the first time we
also observe a visible effect on the scattering observables, when we
introduce an energy shift by setting $E^i=-8$~MeV (dash-dotted line).

\hspace*{10mm}
From Fig.~6 we see that he impulse approximation, even if accurately and
completely calculated, is clearly inadequate in describing the elastic
scattering observables at such a low energy, and additional effects have
to be included. The first term in the spectator expansion contains as
additional term the coupling of the struck target nucleon to the
residual nucleus \cite{med2}. For comparison we include a calculation in
Fig.~6 as dotted line, where this term is calculated in an approximate
fashion as described in Ref.~\cite{med1}. We see that this additional
`medium' contribution is necessary to get a better description of the
spin observables.

\section{Conclusion}

\hspace*{10mm}
We have calculated the full-folding integral for the first-order
optical potential using the impulse approximation within the framework
of the spectator expansion of multiple scattering theory. The exact 
calculation of this full-folding integral requires a three-dimensional
integration, in which the integration variable is coupled to the energy
of propagation of the projectile and target nucleon. We have carried out 
the calculation taking into account the pole structure of the NN t-matrix
when integrating to negative energies. Our optical potentials are based
on a Dirac-Hartree model for the nuclear density matrix and the 
charge-dependent Bonn potential for the NN t-matrix. Recoil and 
frame transformation factors are implemented in the calculation in their
complete form.
We calculate elastic scattering observables for
$^{16}$O,$^{40}$Ca, and $^{208}$Pb at projectile energies from 65 to 
200 MeV laboratory energy and compare the full calculation with
calculations in which the energy of the NN t-matrix is fixed at half
the projectile energy. We find that this fixed energy prescription
describes the full calculation remarkably well for proton scattering at 
200 MeV projectile energy. This leads to the conclusion that the pole
structure of the NN t-matrix does not play a role at intermediate
energies. For projectile energies below 200 MeV we find that the 
influence of the deuteron and di-proton state slowly gains importance as we
approach lower energies. However, between 100 and 200 MeV these effects 
are still relatively small. This is consistent with the small size
of the experimentally measured cross sections for the (p,d) reaction.
At 65 MeV we see considerable differences between the full
calculation and the one in which the energy is kept fixed. In general,
we find smaller differences between our full calculation and the fixed
energy calculations as were suggested in Ref.~{\cite{hugo1}}.

\hspace*{10mm}
The calculations presented in this manuscript are strictly based
on the impulse approximation using the free NN t-matrix.
It has been stated in many places that the impulse approximation is 
insufficient at energies around 100 MeV projectile energy. We confirm
this result, while carrying out complete calculations of the optical
potential taking into account the energy dependence of the NN t-matrix.
Additional corrections have to be taken into account at lower energies.
Within the framework of the spectator expansion these corrections have been
derived and carried out in an approximate fashion \cite{med2}.
A similarly complete calculation, as given here for the first order term
in the impulse approximation, will require a complete Faddeev calculation
of the correction term, which is at present not  possible.

\vfill
\acknowledgments
The authors want to thank W. Gl\"ockle for many stimulating, helpful and
critical discussion during this project.

This work was performed in part under the auspices of the U.~S.
Department of Energy under contract No. DE-FG02-93ER40756 with
Ohio University. We thank the Ohio Supercomputer Center (OSC) for
the use of their facilities under Grant  
No.~PHS206 
as well as the National Energy Research Supercomputer Center
(NERSC) for the use of their facilities 
under the FY1997 Massively Parallel Processing Access Program.



\newpage

\noindent
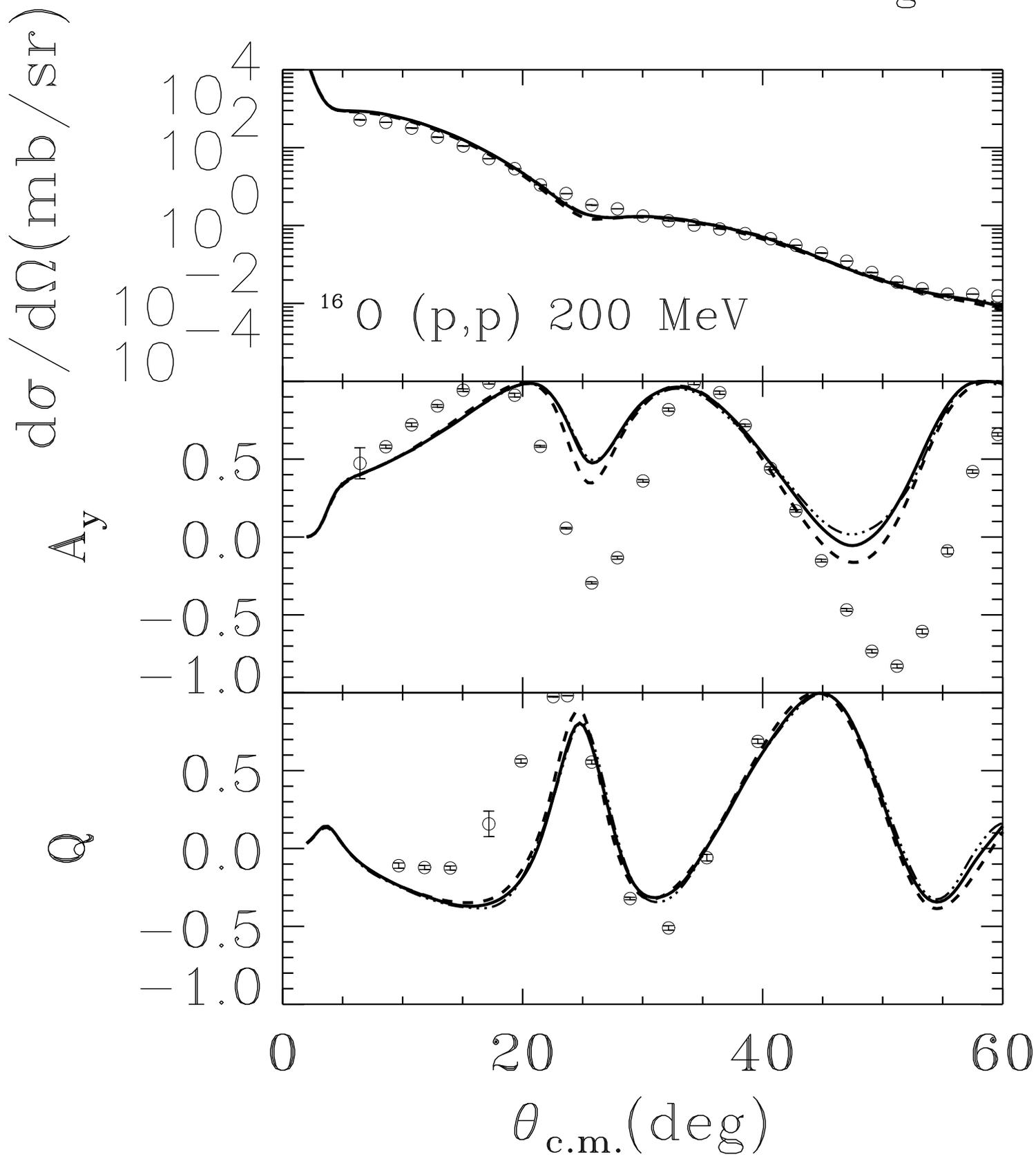
\begin{figure}
\caption{ The angular distribution of the differential cross-section
         ($\frac{d\sigma}{d\Omega }$), analyzing power ($A_y$) and
         spin rotation function ($Q$) are shown for elastic proton
         scattering from $^{16}$O at 200 MeV laboratory energy.
The solid line represents the calculation performed with a
first-order full-folding optical potential based on the DH density
\protect\cite{DH} and the CD-Bonn model \protect\cite{CDBonn} including
the energy dependence of the NN t-matrix. The dashed line represents a
calculation where the energy of the NN t-matrix is fixed at half the
projectile energy. The dash-dotted line stands for a calculation
including the energy dependence of the NN t-matrix and an additional
energy shift by $E^i=-8$~MeV. The data are taken from
Ref.~\protect\cite{O200}. \label{fig1}}
\end{figure}

\noindent
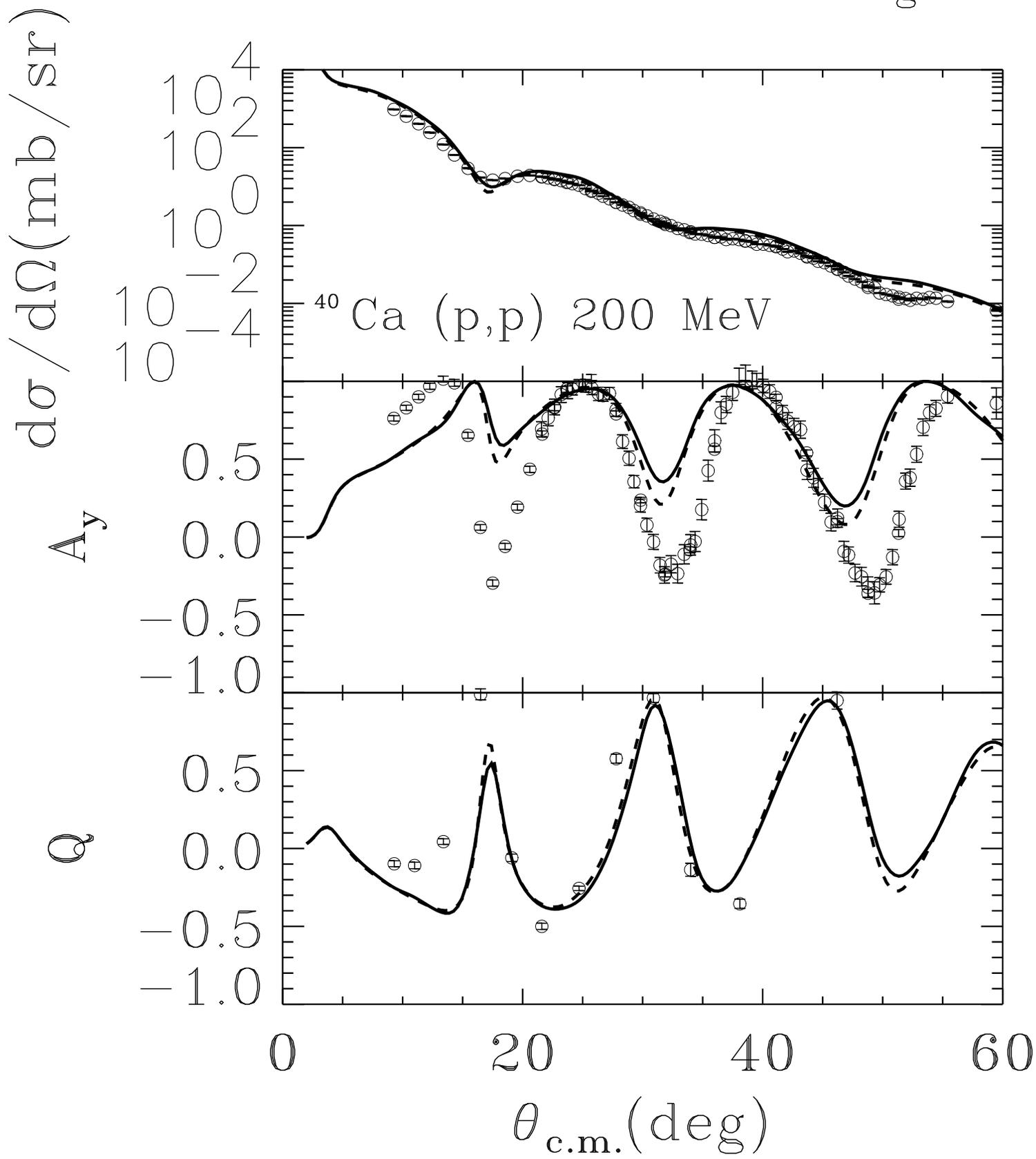
\begin{figure}
\caption{Same as Fig.~1, except that the target nucleus is $^{40}$Ca,
and the dash-dotted line is omitted.
The data are taken from Ref.~\protect\cite{ca200}. \label{fig2}}
\end{figure}

\noindent
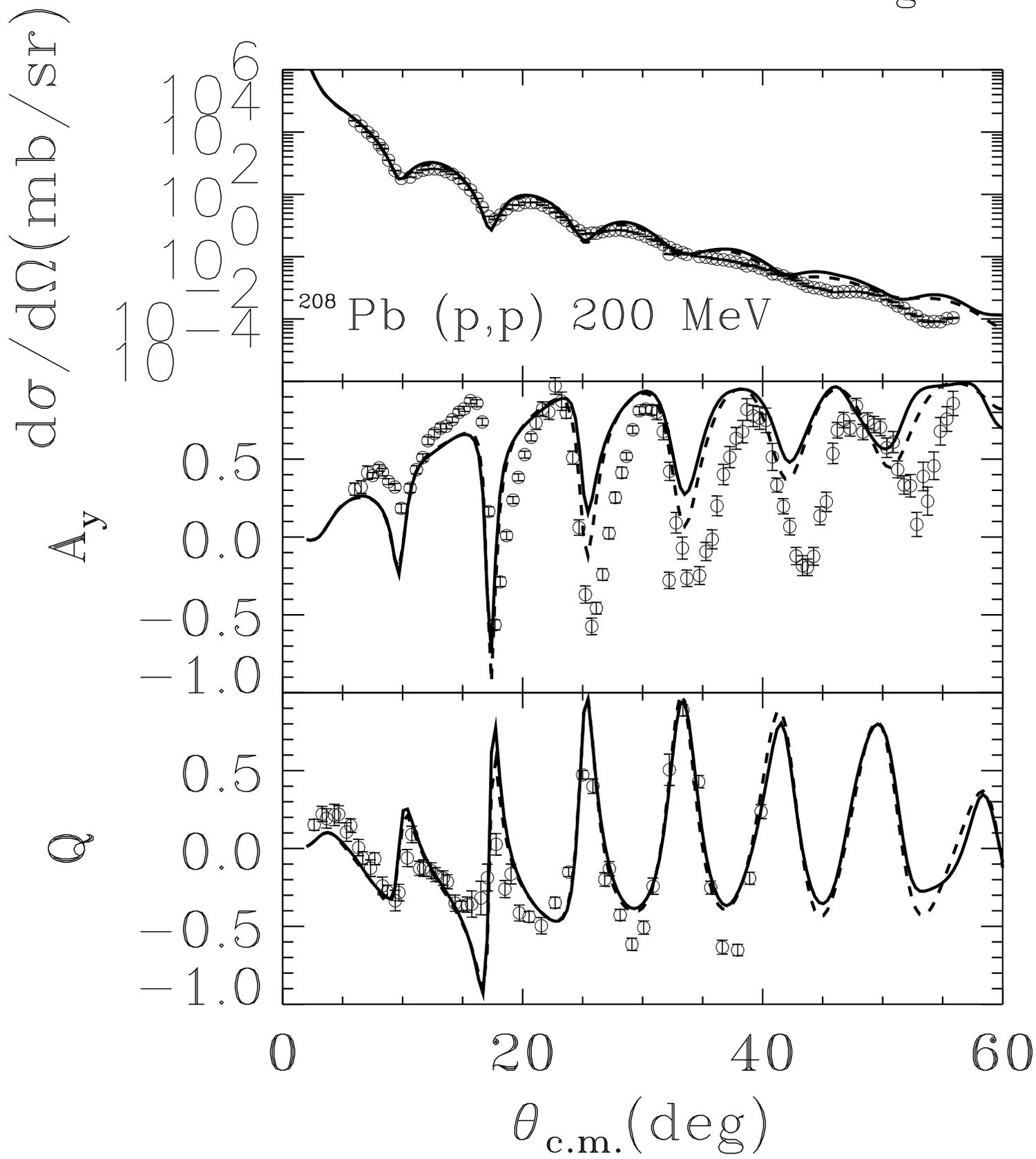
\begin{figure}
\caption{Same as Fig.~2, except that the target nucleus is 
$^{208}$Pb. The data are taken from Ref.~\protect\cite{ca200}. \label{fig3}}
\end{figure}

\noindent
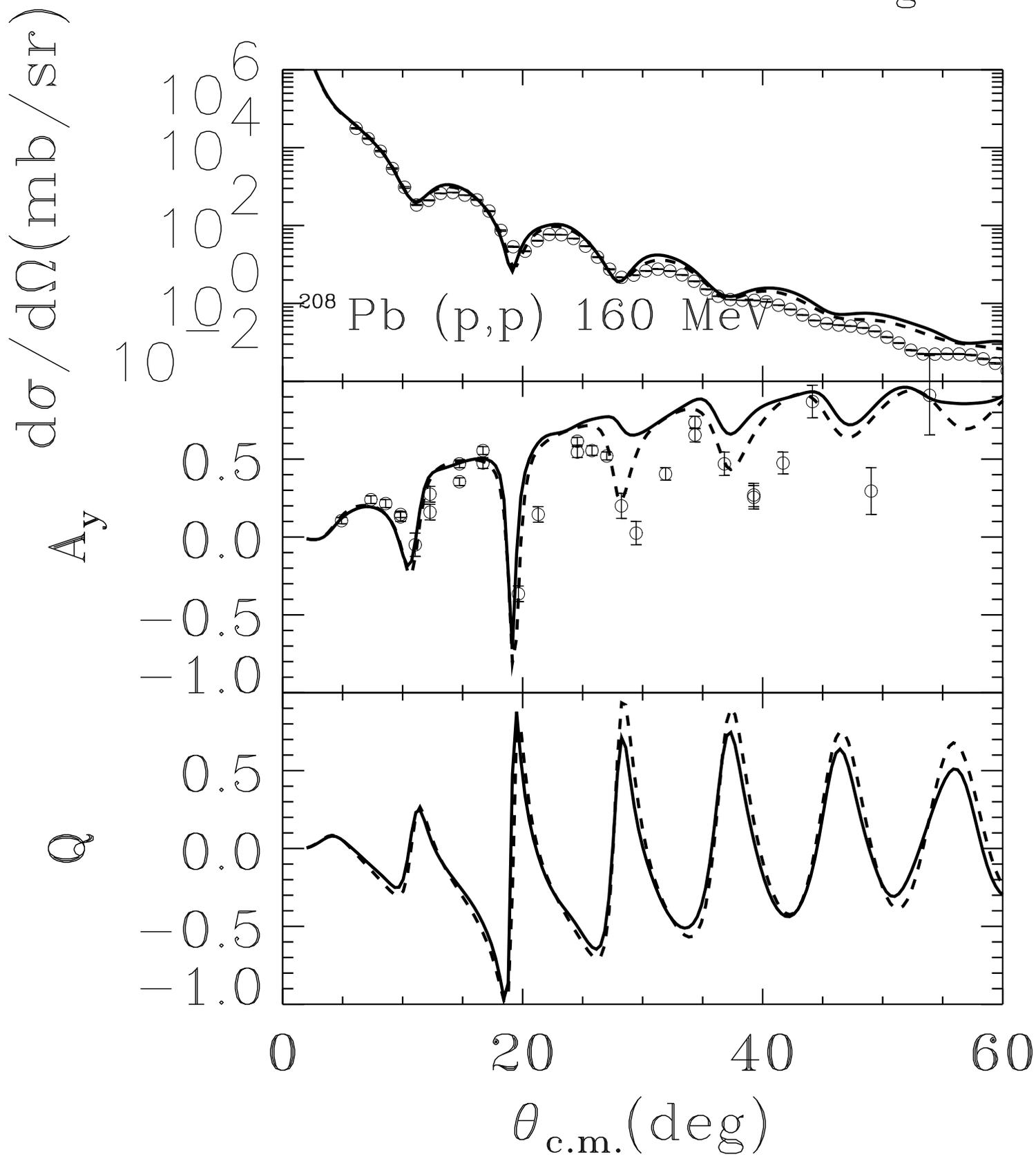
\begin{figure}
\caption{Same as Fig.~3, except for $^{208}$Pb at 160~MeV proton kinetic
energy. The data are taken from  Ref.~\protect\cite{pb160}.
\label{fig4}}
\end{figure}

\noindent
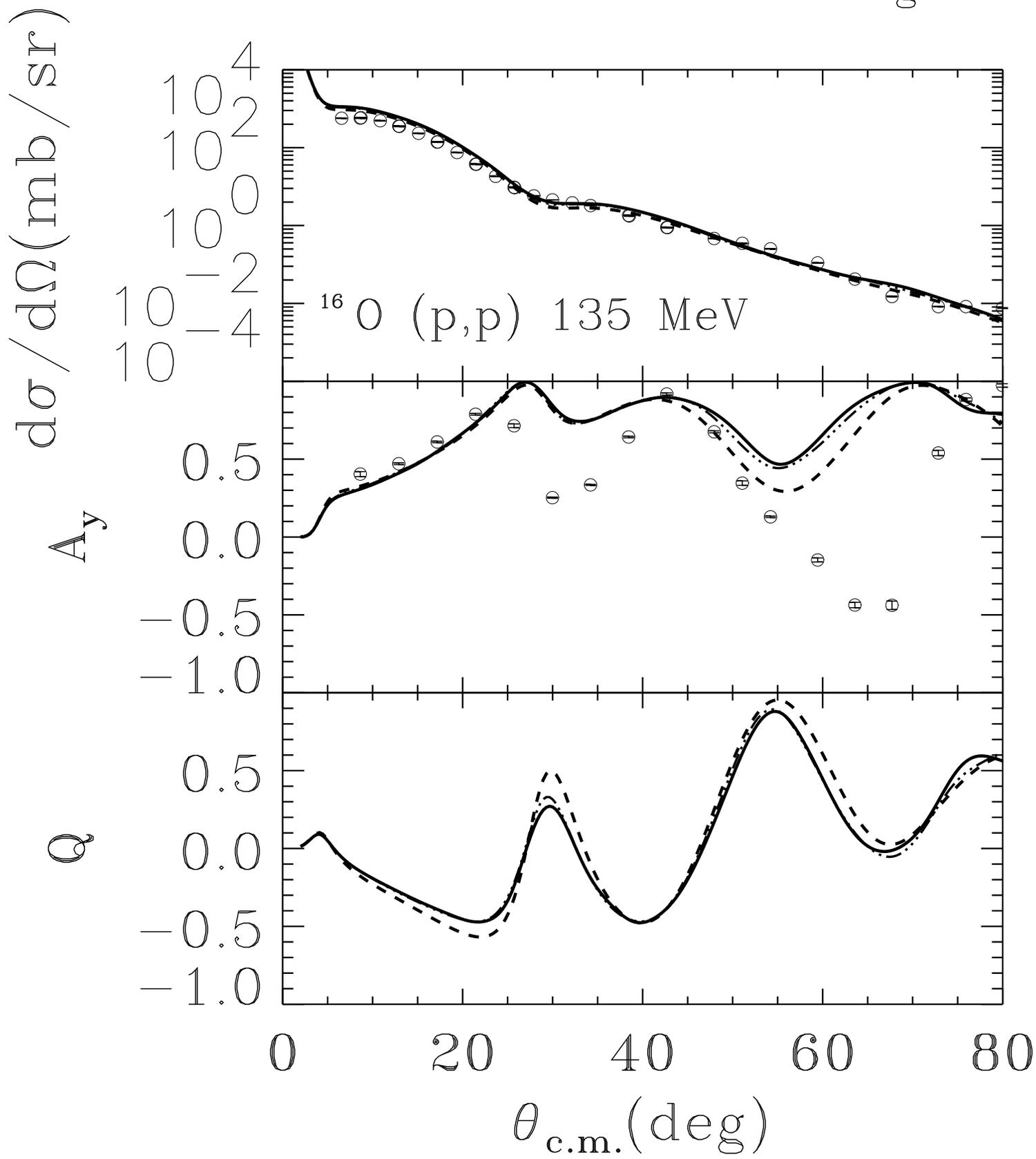
\begin{figure}
\caption{Same as Fig.~1, except for $^{16}$O at 135~MeV proton kinetic
energy. The data are taken from  Ref.~\protect\cite{o135}. \label{fig5}}
\end{figure}

\noindent
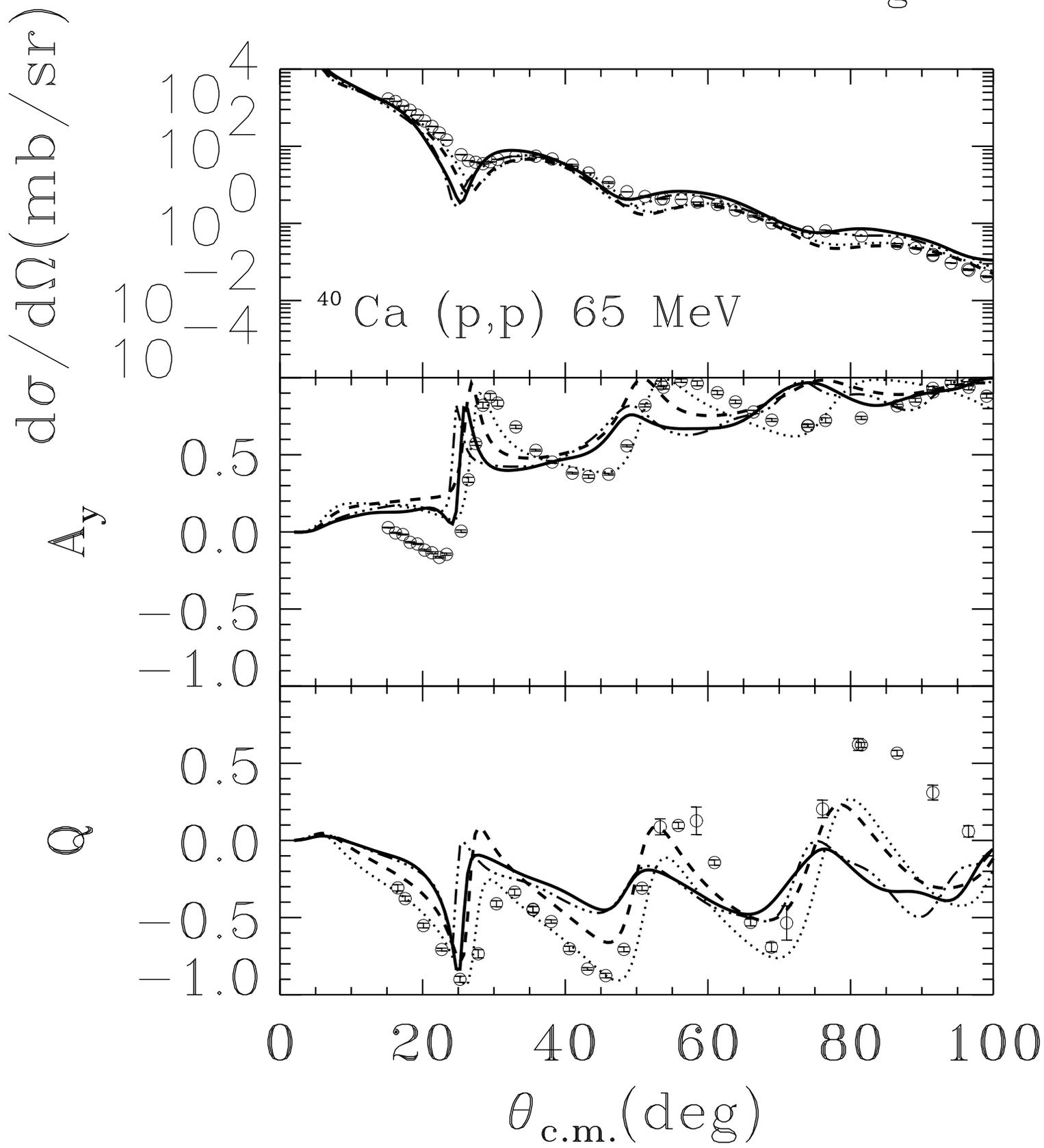
\begin{figure}
\caption{Same as Fig.~1, except for $^{40}$Ca at 65~MeV proton kinetic
energy. The additional dotted line denotes a calculation including the
effect of the coupling of the struck target nucleon to the residual
nucleus. The data are taken from  Ref.~\protect\cite{zr65}. \label{fig6}}
\end{figure}

\end{document}